%% file: main.tex
\documentclass{article}

\usepackage{arxiv}

\usepackage[utf8]{inputenc} 
\usepackage[T1]{fontenc}    
\usepackage{hyperref}       
\usepackage{url}            
\usepackage{booktabs}       
\usepackage{amsfonts}       
\usepackage{nicefrac}       
\usepackage{microtype}      
\usepackage{lipsum}
\usepackage{subfigure}
\usepackage[ruled,vlined]{algorithm2e}
\usepackage{amsmath}
\usepackage{graphicx}
\usepackage{authblk}

\title{University Operations During a Pandemic: A Flexible Decision Analysis Toolkit}

\author[1]{Himanshu Kharkwal}
\author[1]{Dakota Olson}
\author[2]{Jiali Huang}
\author[1]{Abhiraj Mohan}
\author[2]{Ankur Mani}
\author[1]{Jaideep Srivastava}
\affil[1]{Department of Computer Science, University of Minnesota, {\fontfamily{qcr}\selectfont\{khark007 | olso8022 |  mohan056 | srivasta\}@umn.edu}}
\affil[2]{Department of Industrial and Systems Engineering, University of Minnesota, {\fontfamily{qcr}\selectfont\{huan1106 | amani\}@umn.edu}}

\begin{document}
\maketitle

\begin{abstract}
\label{abstract} 
\input{abstract.tex}
\end{abstract}


\section{Introduction}
\label{introduction} 
\input{introduction.tex}

\subsection{Modeling Community Transmission on Campus}
\label{problem}
\input{problem.tex}

\section{Simulation Framework}
\label{simulation}

\input{simulation.tex}

\section{Experimental Evaluation of Policy choices}
\label{exp_eval} 
\input{exp_eval.tex}

\section{Case Study of a Big University Campus}
\label{case_study}
\input{case_study.tex}

\section{Conclusion}
\label{conclusion} 
\input{conclusion.tex}

\bibliographystyle{unsrt} 
\bibliography{refer} 

\end{document}

%% file: abstract.tex
Modeling infection spread during pandemics is not new, with models using past data to tune simulation parameters for predictions. These help understand the healthcare burden posed by a pandemic and respond accordingly. However, the problem of how college/university campuses should function during a pandemic is new for the following reasons: (i) social contact in colleges are structured and can be engineered for chosen objectives, (ii) the last pandemic to cause such societal disruption was over 100 years ago, when higher education was not a critical part of society, (ii) not much was known about causes of pandemics, and hence effective ways of safe operations were not known, and (iii) today with distance learning, remote operation of an academic institution is possible. As one of the first to address this problem, our approach is unique in presenting a flexible simulation system, containing a suite of model libraries, one for each major component. The system integrates agent based modeling (ABM) and stochastic network approach, and models the interactions among individual entities, e.g., students, instructors, classrooms, residences, etc. in great detail. For each decision to be made, the system can be used to predict the impact of various choices, and thus enable the administrator to make informed decisions. While current approaches are good for infection modeling, they lack accuracy in social contact modeling. Our ABM approach, combined with ideas from Network Science, presents a novel approach to contact modeling. A detailed case study of the University of Minnesota's Sunrise Plan is presented. For each decisions made, its impact was assessed, and results used to get a measure of confidence. We believe this flexible tool can be a valuable asset for various kinds of organizations to assess their infection risks in pandemic-time operations, including middle and high schools, factories, warehouses, and small/medium sized businesses.

%% file: introduction.tex
As the events of 2020 have shown, pandemics due to novel viruses can lead to unimaginable disruption in society \cite{OECD}. The impact has been two-fold; the direct impact of the pandemic on physical health and mortality and the indirect impact of lock-downs and social distancing on mental health  \cite{OECD, unevenjobsrecovery, pfefferbaum2020mental} and the economy \cite{unevenjobsrecovery}. A specific example of a critical societal function facing disruption is higher education, especially for starting freshman for whom an important formative experience, namely of transitioning from home to an independent life, has been severely disrupted \cite{reopeningmodels}.
Higher education in the US contributed an estimated \$528 billion\footnote{https://www.ibisworld.com/industry-statistics/market-size/colleges-universities-united-states/} to the national Gross Domestic Product (GDP) in 2020 \cite{hussar2018projections} and employed roughly 3 million people \cite{NCES}. Disruptions in the education sector have long-term ramifications in terms of an inadequately prepared workforce for the future \cite{educationun}. By mid-March 2020 most colleges and universities across the US either cancelled in-person classes or shifted to remote-only instruction and this mode of instruction may continue for an unknown amount of time. In a recent survey of nearly 3,000 institutions \cite{reopeningmodels}, only 21.3\% said that they are considering fully or primarily in person model for Fall 2020 and beyond. 

Operating a major residential university during a pandemic requires making several decisions, with public health guidance coming from organizations like the Centers for Disease Control (CDC) \footnote{https://www.cdc.gov/coronavirus/2019-ncov/index.html} and state health agencies, e.g. Minnesota Department of Health (MDH) \footnote{https://www.health.state.mn.us/diseases/coronavirus/}. Specific decisions include: (i) whether to wear a mask or not, and of what kind, (ii) how much of physical distancing to maintain, (iii) beyond what enrollment should the class be online, (iv) what type of testing to perform, of whom, and how often, (v) what kind of facility management practices to use, e.g. cleaning, ventilation, etc. Not only is there no history to provide guidance on this, even the medical understanding of the pandemic kept evolving over the period during which the decisions had to be made, making the problem even more difficult to address. Despite this, over 3,000 colleges and universities had to develop plans for opening campuses \cite{reopeningmodels}, essentially with no guidance. Public health and epidemiology models of infection spread \cite{smith2004sir, balcan2010modeling, smith1995transims} use past data to estimate parameters and predict public health outcomes in unstructured population. These are helpful for understanding the healthcare burden posed by a pandemic and responding accordingly; and these models are being used in the current scenario as well \cite{covid2020forecasting, covid2020forecasting2, covidactnow, ferguson2020report, imai2020report, riou2020pattern, kucharski2020early, tindale2020transmission, hellewell2020feasibility, wu2020nowcasting, healthcare2020covid}. {\it However, the problem of how college and university campuses should function in this environment is completely new for at least the following reasons: (i) the spread of an airborne disease on campuses in not well understood; campus interaction is structured and often engineered that is not captured by common epidemilogy models (ii) the last pandemic to cause such societal disruption was over 100 years ago, namely the Spanish flu \cite{gates2020responding}, when higher education was not a critical part of society, (iii) not much was known about causes of pandemics, and hence effective ways of safe operations were not known, and (iv) today we have distance learning, via which remote operation of an academic institution is possible.} Thus, the problem of managing the re-opening and operations of a university during a raging pandemic is only 6 months old; and development of techniques to address it are at a nascent stage. Ever evolving knowledge of the virus, no experience with socially distanced operations, and severe health risk of decisions, has made it even more difficult for administrators \cite{mossa2020policies, sun2020covid}. Details of \emph{disease progression}, \emph{infection transmission}, and \emph{social contact patterns} in response to operational decisions are largely unknown and evolving over time. However, administrators need to make decisions; which creates the need for a tool that can assess the future impact of decisions and provide guidance. Additionally, it needs to be \emph{flexible} and \emph{adaptable} to the changing knowledge base and data. 
In this paper, we present a flexible framework for impact analysis of university operating policies during a pandemic. Our simulation framework includes a library of models, a simulation environment, and a visualization component that allows policymakers to consider a range of policies.

%% file: problem.tex
\begin{figure}[ht]
\centering
\small
\includegraphics[width=13cm,height=6cm]{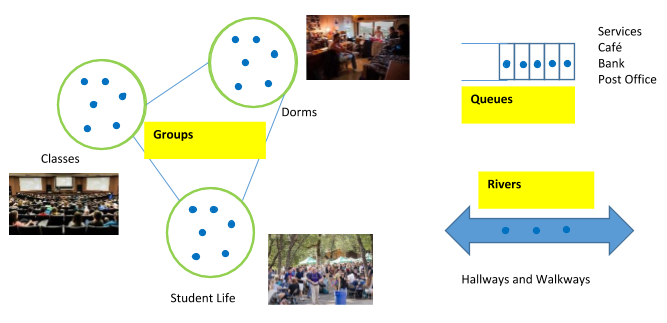}
\caption{Engineered Interactions on Campus}
\label{fig:campus}
\end{figure}

The campus environment has some unique features as compared to other places. Figure \ref{fig:campus} shows types of interactions on university campus, namely {\it groups, queues} and {\it rivers.} Groups can further be classified into on-campus interactions. that can be monitored and controlled and off-campus interactions that cannot be monitored. The former includes classrooms, study areas, student life activities under the university's purview, e.g. dorms, extra-mural sports, clubs, etc. Off-campus activities include private housing, social activities, grocery shopping, and a myriad of other life activities. Queues appear at various kinds of service locations on campus, including those offered by the university, e.g. bookstores, student services, etc., as well as those offered by others, e.g. cafes, banks, etc. Rivers include pathways where students cross each other. This classification of interactions allows for more accurate modeling of disease spread. Given the nature of the pandemic, groups are the most risky types of interactions, involving sufficiently large number of people in close proximity for long periods of time. Further, it is comparatively easier to reduce community transmission in queues using appointments only service and in rivers using one-way rivers, ventilation, masking and physical distancing rules. Thus, among all different types of interactions, the on-campus group interactions are the ones that have high community transmission risk, are observable and controllable. Therefore, it is reasonable to have a sophisticated model for community transmission through on-campus group interactions, in particular classrooms and a simple model of infections through other types of interactions consistent with related work \cite{gressman2020simulating}. In our work, we include a detailed stochastic network model of classes, especially since there is growing evidence of virus transmission through aerosol spread \cite{anderson2020consideration, banik2020evidence}. Since classes have fixed schedules, they can be modeled as processes happening at specific times, with batched arrivals of students and instructors. Classrooms are assumed to be completely cleaned and sanitized between any two consecutive classes thus the transmission is limited within classes and mixing across two classes is only due to the same people in the two classes and not due to the shared classrooms. Infections outside the classroom settings are random with chances of infection depending upon the community prevalence rates \cite{gressman2020simulating}.

\subsection{Candidate Policies for University Operations}
Based on our review of various discussions in the media and other sources over the past 6 months, and confirmed by our interactions with university administration, we identify the following key decisions (dimensions) to be considered:
\begin{enumerate}
    \item Masking: What type of mask and compliance policy.
    \item Physical distancing: Student classroom density, in sqft/student, based on the physical distance between students.
    \item Class modality: In-person or online classes.
    \item Testing: Testing policy, i.e. symptomatic or asymptomatic, and whether to do contact tracing.
\end{enumerate}

Any operational policy consists of a set of choices, one for each of the decisions outlined above. An example of a policy is Minnesota's Sunrise Plan \cite{umnsunrise}, details of which are presented in Section \ref{case_study}. The University administration needs to evaluate the cost of implementing a policy and the benefits obtained from them. Implementation costs for policies are usually estimated based upon the predictions of behavior. For example, class modality decisions can incur technology, infrastructure and support costs, as well as revenue loss due to changes in student enrollment. Our work provides a method of estimating the benefits of the operational decisions and policies.

\subsection{Our Contributions}
In this paper, we present a flexible framework that allows for various models for human behavior and socialization patterns, decisions and choices, infection transfer models, disease progression models, etc. to simulate the outcomes under different policies. 
A simulation to evaluate policy impact requires several models. For the problem of evaluating University operations policy, four models are key, namely (i) a model of social behavior and interaction patterns among people, (ii) a model of infection transfer from people to people, which can be direct (person-to-person contact) or indirect (contact intermediated by temporally and spatially co-located visits to a location where infected people deposit the infection, and the susceptible people pick up the infection), (iii) a model of disease progression in an individual once infected, and (iv) the management policy being used. Once a set of models are selected from the model library and a management policy is chosen, a scenario needs to be executed to assess the impact. In our simulation framework there are three components that achieve this, namely the Person-Location Visit Generator, the Infection Transfer Generator, and the Disease Progression Generator. Each of these uses the corresponding model selected for the execution. For the infection Transfer Generator and the Disease Progression Generator, several models exist, and we give brief descriptions of one of each kind in Section \ref{sec:infection_transfer} and Section \ref{sec:disease_progression}. For the Person-Location Visit Generator there are no good models and based on our ongoing work we propose to build new ones. These are described in detail in Section \ref{sec:person_location}. 



The framework we developed is flexible since it allows a policy analyst to experiment with various types of models and policies. In addition, it is expandable because new kinds of models, policies, as well as metrics and visualizations can be added. 

\paragraph{New Model of Social Contact}

One important contribution of the paper is a new model of social contact. Disease spread depends upon the pathogen properties as well as the contact structure in the population. We introduce a people-place network model for social interactions that replaces the unstructured population model used for pandemic modeling. The disease spread prediction models for COVID-19 so far have ignored the heterogeneity and randomness in the contact structure of the population \cite{ferguson2020report, covid2020forecasting2, covid2020forecasting, healthcare2020covid}. The models are based upon the variations of compartmental models such as SI, SIR, SIS, SIRS, SEIR \cite{bass1963dynamic, rogers2010diffusion, anderson1991discussion, hethcote1989three, hethcote2000mathematics, strogatz2018nonlinear} that assume populations with homogeneous interactions and give rise to simple ordinary differential equations. The models were originally developed by medical doctors in early twentieth century and later studied by mathematicians, engineers and social scientists. There is limited use of structure in interactions, only at a coarse level in \cite{OECD} based upon \cite{balcan2009multiscale}.

\paragraph{Case Study of a Major University's Policy:} Over summer 2020, we interacted with the University of Minnesota Administration administration to track various decisions, and analyzed a range of choices for each discussion, to help inform the decisions. We believe this case study, in addition to showing the usefulness of our approach, also provides helpful guidance for the future.

\subsection{Outline}
The rest of the paper is organized as follows: Section \ref{simulation} describes the design of the toolkit, Section \ref{exp_eval} provides an evaluation of the choices for each decision and its impact, Section \ref{case_study} provides a detailed case study of the University of Minnesota's Sunrise Plan, and Section \ref{conclusion} concludes the paper, with potential directions for future work.

%% file: simulation.tex
We now present a flexible stochastic simulation framework for evaluating university operational decisions that contains libraries for (i) structural and behavioral models of human contact calibrated using real data, (ii) disease transmission models in buildings, and (iii) disease progression models. Current COVID-19 spread models are strong on modeling infection spread and disease progression, but relatively weak in modeling human behavior and social contact in various campus activities, e.g. attending classes. This leads to estimates that have a high degree of deviance from reality \cite{covidactnow, OECD}. Proposed approach uses an agent based model (ABM) for human interaction, and stochastic models for physical and biological processes. The ABM models human interactions as a network, which provides better predictions of disease spread than traditional SIR and SEIR models, which are population based.
Given the airborne transmission mode of COVID-19, i.e. aerosolized droplets containing the pathogen stay suspended in air for long periods of time \cite{anderson2020consideration, van2020aerosol}, indirect human contact must be modeled in addition to direct contact. Therefore, we introduce a \emph{Person-Location Network}, a bipartite graph that captures human interaction indirectly though locations visited.

\begin{figure}
\centering
\small
\includegraphics[width=15cm,height=7cm]{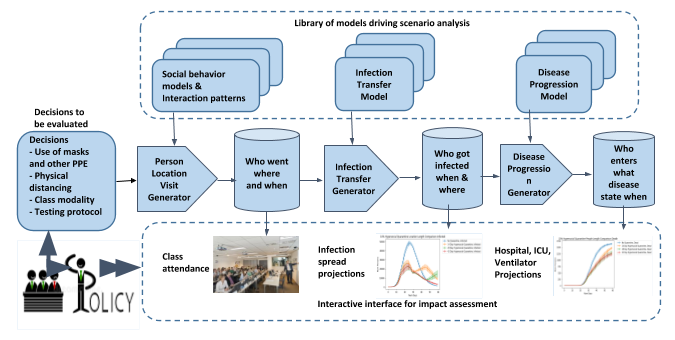}
\caption{System Architecture}
\label{fig:arch}
\end{figure}

Figure \ref{fig:arch} shows the architecture of the simulation framework, consisting of three system components: the Person-Location Visit Generator, the Infection Transfer Generator, and the Disease Progression Generator. There exist several models for each of these components developed at different times as the knowledge about the disease evolved, along with available data such as list of courses for Fall 2020, course selections, mask use policy, number of in person courses, and number of students, faculty, and staff on campus. In the following, we describe the latest models which we have implemented for each component.

\subsection{Person-Location Visit Generator}\label{sec:person_location}
The Person-Location Visit Generator creates a sequence of visits, i.e. events of individuals visiting locations. In the following, we define the network and describe the generation process of the network and the event sequence. 

\paragraph{Person-Location Bipartite Network}
The basis of the Person Location Visit Generator is the person-location bipartite network $G(P,L)$, with $P$ and $L$ the set of people and locations, respectively. This network captures all connections between people and locations, i.e.
existence of an edge between $i$ and $j$ means that person $i$ visits location $j$ at some point. In the simulation, this network $G(P, L)$ is a realization from some random network generation process.

\paragraph{Network Generation:}
To generate the network $G(P, L)$, we consider the case where there are $N$ people and $M$ locations in the bipartite network, i.e. $|P| = N$ and $|L| = M$. Input to the generation process includes the set of nodes $P$ and $L$, degree sequence of nodes in $P$, denoted as $d_1, \cdots, d_N$, and the degree sequence of nodes in $L$, denoted as $w_1, \cdots, w_M$. These two degree sequences can either be obtained from data, or generated as random samples from certain degree distributions. Note that degree distributions for the people side and the location side can be different. Also, we may need additional adjustments of the two degree sequences to ensure $\sum_{i=1}^N d_i = \sum_{j=1}^M w_j$, so that they are valid sequences for the bipartite network. 

The bipartite network is then generated as a realization from the configuration model \cite{bender1978asymptotic, molloy1995critical} with given set of nodes $P$ and $L$, and desired degree sequences $d_1, \cdots, d_N$ (for nodes in $P$) and $w_1, \cdots, w_M$ (for nodes in $L$). The pseudo code is highlighted in Algorithm \ref{lgm:configuration}. This algorithm returns the person-location bipartite network $G(P, L)$.

\begin{algorithm}
\SetAlgoLined
\SetKwInput{KwInput}{Input}
\SetKwInput{KwOutput}{Output}
\SetKwInOut{KwInit}{Initialization}
\caption{Network Generation Process}
\KwInput{Set of nodes $P$ and $L$; degree sequences $d_1, \cdots, d_N$ (for nodes in $P$) and $w_1, \cdots, w_M$ (for nodes in $L$).}
\KwOutput{Return the generated person-location bipartite network $G(P,L)$.}
\KwInit{$S = \sum_{i=1}^N d_i = \sum_{j=1}^M w_j$.}
Assign $d_i$ half-edges for person $i$, $i = 1, \cdots, N$ and $w_i$ half-edges for location $j$, $j = 1, \cdots, M$\;
\While{$S>0$}{
  Choose one half-edge from the people side and one half-edge from the location side, both uniformly at random across all half-edges on the same side\;
  Connect the two half-edges to form an edge, and add it to the edge set of $G(P, L)$\;
  S = S-1.
  }
 \label{lgm:configuration}
\end{algorithm}

\paragraph{Event sequence:}
The input data for our simulation is an event sequence $G(P, L, T)$, where $T$ is the discretized time range for the whole simulation. For each time $t\in T$, $G(P, L, t)$ represents the actual visits between people and locations at time $t$ and is a subgraph of $G(P,L)$. We can see that $G(P, L, t)$ is still a bipartite network.

$G(P, L, T)$ can be obtained from data or generated as random samples. One simple way to create $G(P, L, T)$ as random samples can be: for each time $t$, we sample the set of edges in $G(P,L)$ with probability $p$ uniformly at random, and denote the resulting subgraph as $G(P,L,t)$. The parameter $p$ captures the sociability of people and locations. Within a region, certain areas may have higher sociability factor $p$, while other areas may have lower sociability factor $p$. Many factors, such as user behaviors, regional characteristics and customs, and geographic and weather conditions, can be integrated into this parameter $p$. More importantly, $p$ can also be modified to capture the impact of some public policies due to the outbreak of COVID-19. For example, shutdown or reduced operations of business as well as shelter-in-place can be modeled as reducing the value of the sociability factor $p$.


\subsubsection{Modelling a Campus}
\label{sec:campus_generator}
Since we are particularly interested in the use case of university re-opening, we will present the campus specific Person-Location Visit Generator in this section. As discussed in Section \ref{problem}, by assumption, we will only model the interactions within classes. 

\begin{figure}
\centering
\small
\includegraphics[width=8cm]{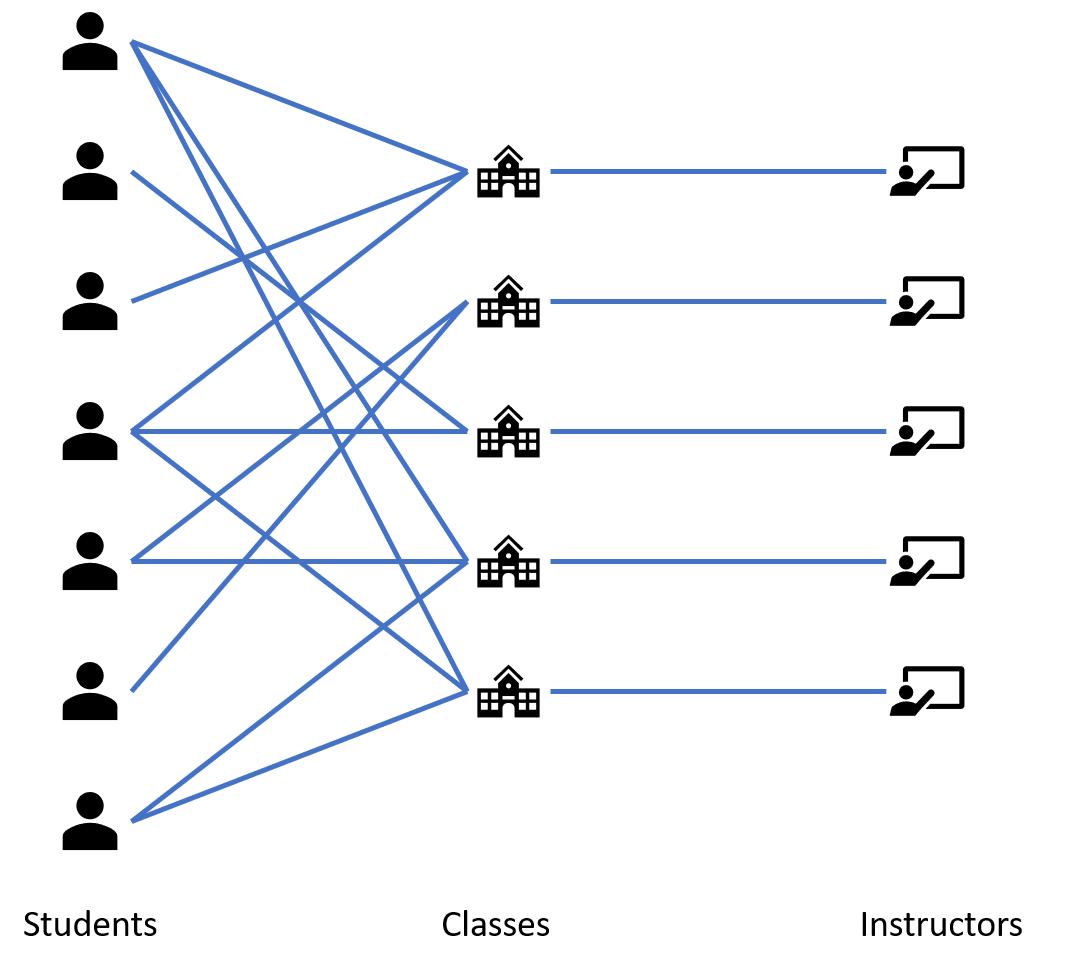}
\caption{People (Students and Instructors) and Locations (Classes) bipartite network}
\label{bipartite_network}
\end{figure}

In modelling a campus, we generate a student / instructor-class bipartite network $G(S\cup I, C)$, where $S$ denotes the set of students, $I$ denotes the set of instructors, and $C$ denotes the set of classes. $G(S\cup I, C)$ models which student is taking which class as well as which instructor is teaching which class. Each student has a student profile indicating their department and academic level. Similarly, each class has a class profile indicating its department and difficulty level. Each instructor is assigned to exactly one class and this forms the instructor-class part of the bipartite network. The student-class part of the bipartite network is generated by a modified configuration model, where students are assigned to classes following certain restrictions. We assume that each student chooses $2$ to $5$ classes (this forms the degree sequence of the students) subject to the capacities of classes (this forms the degree sequence of the classes). With probability $p_1$, students choose classes within their own department with difficulty levels matching their academic levels; with probability $p_2$, students choose classes within their own department with difficulty levels not matching their academic levels; with probability $p_3$, students choose classes outside their own department. In general, $p_1 > p_2 > p_3$ and we also require $p_1 + p_2 + p_3 = 1$. These restrictions can be imposed by integrating networks from multiple configuration model processes. We point that generating the network $G(S\cup I,C)$ from such random processes is useful for analysis prior to course enrollment; if we have the exact student-class enrollment data and instructor-class data, then we can create a deterministic bipartite network $G(S\cup I, C)$ from the data. Figure \ref{bipartite_network} shows the topology of the bipartite network $G(S\cup I, C)$.

Based on the bipartite network $G(S\cup I, C)$, and the teaching schedule of all classes throughout the semester, a visit schedule of who (students and instructors) visits which class and when will be created. If we assume that students will always attend classes, then this visit schedule becomes the natural event sequence $G(S\cup I, C, T)$ for the simulation, where $T$ denotes the set of days within the simulated academic semester. Otherwise, we can introduce an attendance rate $p$ to generate the event sequence. The attendance rate $p$ captures the students' attendance activity: with probability $p$, a student will attend a scheduled class; with probability $1-p$, a student will skip a scheduled class. For day $t$, we sample the set of edges in the visit schedule associated with that day with probability $p$ uniformly at random, and denote the resulting graph as $G(S\cup I, C, t)$. The event sequence $G(S\cup I, C, T)$ can be obtained as $\{G(S\cup I, C, t): t \in T\}$.  


\subsection{Infection Transfer Generator}\label{sec:infection_transfer} 

The Infection Transfer Generator generates a sequence of infections using the event sequence generated by the Person-Location Visit Generator and a disease transmission model. The probability of infection is computed using the Wells–Riley equation \cite{noakes2006modelling, dai2020association} commonly used for modeling indoor airborne disease transmission. The Wells-Riley equation is used to calculate the probability of each susceptible individual of getting infected when indoors (as in classrooms) with other infectious people. The probability of infection depends upon the physical environment, including room volume, ventilation rate, time spent in the room, pulmonary ventilation rate, and the infectiousness of the disease (quanta of pathogen). The probability calculation by the Wells-Riley equation is given by the following:
\begin{equation}\label{eqn:wells-riley}
   P = \frac{C}{S} = 1-e^{-\frac{Ipqt}{Q}},
\end{equation}
where $P$ is probability of infection, $C$ is the number of newly infected people, $S$ is the number of susceptible people, $I$ is the number of infectors, $p$ is the pulmonary ventilation rate of susceptible (${m^3}/h$), $Q$ is the room ventilation rate (${m^3}/h$), $q$ is the quantum generation rate (quanta$/h$), and $t$ is the exposure time. The equation demonstrates how changes to both the physical environment and infection control procedures may potentially impact the spread of airborne infections in indoor environments such as classrooms.

The original Wells-Riley equation is used for fast-moving infections. It assumes that during the scenario that a group of people are in an indoor environment, there is a chance that susceptible individuals exposed to the pathogen produced by infected individuals may get infected and start adding pathogen to the environment. This is unlikely in classroom settings because the class time is much small than the usual incubation period of COVID-19. Alternatively, the disease quantum generation rate is very small, making the infection transfer within a classroom slow-moving. We introduce a simple solution to this problem. We use the first-order Taylor approximation of the Wells-Riley equation \eqref{eqn:wells-riley} to model the transmission probability. We ignore the higher-order terms in the Taylor series that are not valid because of the high incubation time and low quantum generation rate of COVID-19. In particular, we are using the following equation to model the probability:
\begin{equation}\label{eqn:wells-riley taylor approx}
    P = \frac{Ipqt}{Q}
\end{equation}
This approximation of the Wells-Riley equation is extremely close to the original equation because the exposure time in classrooms is small and a linear approximation of an exponential function is very accurate when the argument of the function is small.

\subsection{Disease Progression Generator}\label{sec:disease_progression}

\begin{figure}[ht]
\centering
\includegraphics[width=0.7\textwidth]{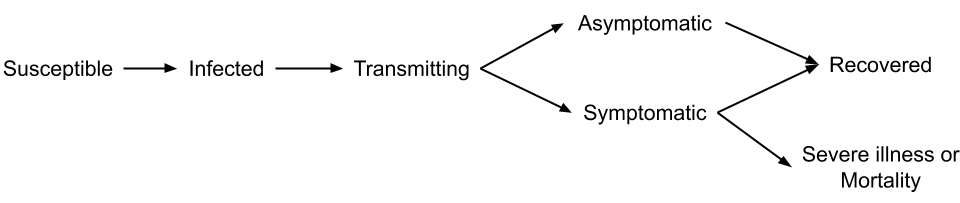}
\caption{Disease progression stages}
\label{fig:progression}
\end{figure}

The disease progression generator library generates the disease state transitions of each agent. The key states of the library include Susceptible, Infected, Transmitting/Infectious, Asymptomatic, Symptomatic, Severely ill, Dead, and Recovered, which are highlighted in Figure \ref{fig:progression}. It shows a standard finite state epidemiological model of disease progression states \cite{smith2004sir}, where arrows imply the direction of change of state. Our simulator follows this epidemiological model and obtains the distribution of time spent at different states from existing literature.

The transition from the susceptible to the infected state primarily occurs according to the Infection Transfer Generator and is probabilistically determined by the first-order Taylor approximation of the Wells-Riley equation \eqref{eqn:wells-riley taylor approx} inside the campus, Secondly, it also happens spontaneously through outside infection, whose rate is a variable that depends upon the disease prevalence around the campus. Once infected, the agent starts his / her incubation period, which follows a Weibull distribution $Weibull(0.11, 1.97)$ with a mean of 8.29 days and a median of 7.76 days, based on \cite{qin2020estimation}. People become infectious and start transmitting viruses 1-3 days (uniformly distributed) before their incubation period ends \cite{he2020temporal}. Once an agent starts transmitting, it becomes asymptomatic (this includes agents that are pre-symptomatic).

When the incubation period ends, the agent either remains asymptomatic or transitions to being symptomatic with 65\% probability \cite{asymrateCDC}. We assume that the symptomatic or asymptomatic agents remain contagious until recovered. The distribution of this period has a mean of 7.8 days and is best fit by the gamma function $Gamma(3, 7.8/3)$ \cite{UMNPH}. These are parametric values that can be modified as we change our understanding of the disease or implement different testing policies. 

Asymptomatic or symptomatic agents with incorrect test results (False Negative rate is assumed to be 3.3\% of total tested based on the test mentioned in section \ref{sec:individual-testing}) go around spreading the disease to locations they visit and eventually will transition to the recovered state at the end of their contagious period \cite{Wajnberg}.
Agents with positive test results are taken out of the simulation and put in effective quarantine based on the number of days left in their contagious period. A portion of them, defined by a probabilistic parameter, will develop a severe illness or even mortality and will not be able to come back on campus throughout the semester. For those without severe illness or mortality, once the quarantine period ends for them, they are put back into the simulation as recovered state, which entails that state for the remainder of the semester and thus they will not get infected again. We have based this on the informed assumption that antibody immunity lasts for three months which is more than the entirety of our semester \cite{vanKampen}.

We point out that our Disease Progression Generator library is general enough to model the complete disease progression. For example, we can also include states like hospitalized, shortness of breath, respirator, ICU, dead, etc. into the library and perform further analysis. However, controlling the spread of the disease is the primary concern for universities, therefore we omit those states and focus more on the infection and transmission of the disease. For this study, we analyze the cumulative infected students due to community transmission of COVID-19 in section \ref{exp_eval}, hence the fraction of agents who leave the system (severe illness or mortality) or get recovered is immaterial for our simulations because neither of the states impact new infections. Recovered patients are immune and don't act as vectors whereas patients who leave are isolated from the system.

%% file: exp_eval.tex
The simulator, available in Github\footnote{https://github.com/shepherd13/Covid-19\_university\_reopening\_framework.git}, was used to evaluate the effectiveness of different operational interventions for reducing infections during the semester. The University of Minnesota is chosen, to represent many big universities in the country. Following describes the data and assumptions used for the experiment in Section \ref{sec:dataset_simulation} and Section \ref{sec:parameter_choises} and analyzes the impact of different operational interventions in Section \ref{sec:analysis_policy_dimensions}. 

\subsection{Dataset}\label{sec:dataset_simulation}
The simulator uses actual student class enrollment data from the University of Minnesota and contains 46,782 students, 5,570 classes and 5,570 instructors. These are from Fall 2019  Official Enrollment Statistics Report \footnote{https://oir.umn.edu/student/enrollment}. We consider students from each department except College of Continuing and Professional Studies, since students from that college primarily take online courses. Thus, our event sequence is $G(S\cup I, C, T)$, where $|S| = 46782$, $|I| = 5570$, $|C| = 5570$, and $|T| = 7 \times 12 = 84$ days. 
Each student can enroll in 2 to 5 courses as per university guidelines to maintain student status, which defines the degree of each student in the network $G(S\cup I, C)$. We assume that students can only take classes within departments of their own school / college in the results we present. Fall 2019 registration information of UMN \footnote{https://dept.aem.umn.edu/cgi-bin/courses/noauth/class-schedule} is used to provide the set of classes offered, number of students enrolled in each class, timings, and the instructors to the simulation. With such information, we construct the person-location bipartite network and event sequence according to Section \ref{sec:person_location}.

\subsection{Parameter choices}\label{sec:parameter_choises}
Although the current focus is on the pandemic operations of a major university, the framework is flexible enough to analyze the  spread of infectious diseases involving human interactions in a big campus if any kind, given relevant models and parameters. As we develop the understanding of the disease, we regularly update the parameters involved in the framework based on the current studies. Below we discuss some of the major parameters used in the framework.

\paragraph{Initial infection:} We start our simulation with a sample of the population being initially infected. For our analysis, we assume this value to be 1\% according to the Minnesota's weekly COVID-19 reports \cite{weeklycovidreport}. We randomly select the initially infected people in the simulation, and uniformly distribute them into different groups based on the number of days since they have been infected (maximum 5 days). 

\paragraph{Outside transmission:} Students lead a significant portion of their life outside the university and this is not precisely modeled in the simulation, especially since it cannot be controlled. Further, 75\% - 80\% of students live in off-campus private housing, details of which are outside of the university's purview. Detailed modeling of this has not been done. Instead, an assumption is made that 5 non-quarantined susceptible students are spontaneously infected every day due to presumed transmission from non-university contact. This is consistent with the analysis by the UPenn/Swarthmore team \cite{gressman2020simulating}. We expect modelers to have better estimate of this parameter as the semester progresses. 

\paragraph{Indoor transmission:} We consider indoor transmission of COVID-19 to be airborne, based on the assumption that tables, chairs, equipment and other surfaces inside the classrooms are being systematically sanitized by the university cleaning staff. We use an approximation of the Wells-Riley equation \eqref{eqn:wells-riley taylor approx}, which already has well defined parameters for airborne transmission inside a classroom. The pulmonary ventilation rate of susceptible, defined as $p$, is set at $0.48\  {m^3} /h$ \cite{noakes2006modelling}. The quantum generation rate $q$, or the amount of infection produced by a COVID-19 patient per hour, is assumed to be 20 quanta$/h$ \cite{dai2020association}. We also assume that different types of masks have various efficiency 
in filtering the quanta generated (COVID-19 virus released) and pulmonary ventilation (air intake). This is presented in the detailed analysis of masking impact in Section \ref{sec:analysis_policy_dimensions}. The room ventilation rate $Q$ is assumed to be the product of ventilation rate $4\  ac/h$ \cite{noakes2006modelling} and room volume (explained in Section \ref{sec:Physical_distancing}). 

Time period spent by each agent at a disease progression state is obtained from appropriate continuous probability distribution. The distributions and parameters pertinent to each state have already been discussed in section \ref{sec:disease_progression}. Parameters related to each policy dimension are explained in Section \ref{sec:analysis_policy_dimensions} in detail, along with their individual analysis.

\subsection{Analysis of Individual policy dimensions of the Model} \label{sec:analysis_policy_dimensions}
In this section we analyze the effects of individual policy dimensions, including masking, physical distancing, class modality, and testing. Specifically, we estimate the cumulative infected students due to the community transmission of COVID-19 within the university campus. To analyze the impact of each policy dimension, we vary values of parameters associated with that policy dimension, keeping the other dimensions as low as possible (i.e., equivalent to pre-COVID-19 times). For each set of simulation parameters we ran 1,000 simulations, plot the mean of cumulative infected students and the 95\% confidence interval for the mean. Across different simulations the results are concentrated and the standard deviations are relatively small as compared to the mean values. Thus, the 95\% confidence intervals are close to the mean values.  

\subsubsection{Masking}

The masking policy is an aggregation of four parameters, namely, Student Mask Type, Student Mask Compliance, Instructor Mask Type, and Instructor Mask Compliance. We study the impact of this policy from two aspects, namely the Student Mask Compliance (the percentage of student population that wear a mask) and the Student Mask Types. For the study of Student Mask Compliance, students and instructors are assumed to be wearing cloth masks. The Instructor Mask Compliance is set to 100\%, while the Student Mask Compliance is set as a variable. For the study of Student Mask Types, we consider three types of masks, namely cloth masks, medical masks, and N95 masks. The effectiveness of different mask types was modeled based on a study in \cite{jung2013comparison}, comparing the filtration efficiency of small aerosols on different types of masks. The study \cite{jung2013comparison} found that, on average, N95 masks filter efficiency was 95\%, medical mask filter efficiency was 55\%, and general cloth mask filter efficiency was 38\%. We also assume that all classes are in-person, physical distancing is 2 feet, and only symptomatic people are tested.

\begin{figure}
\centering
\includegraphics[width=\textwidth]{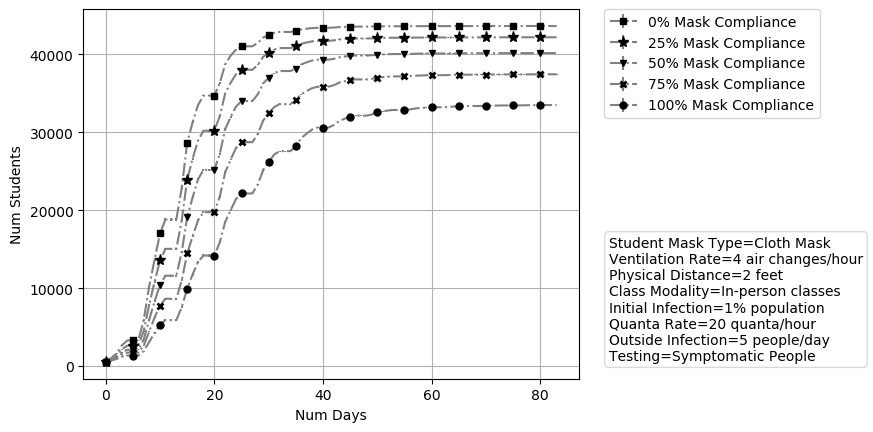}
\caption{Impact of different mask compliances on cumulative infected students due to the community transmission of COVID-19 within university campus}
\label{fig:individual-masking1}
\end{figure}

\begin{figure}[ht]
\centering
\includegraphics[width=\textwidth]{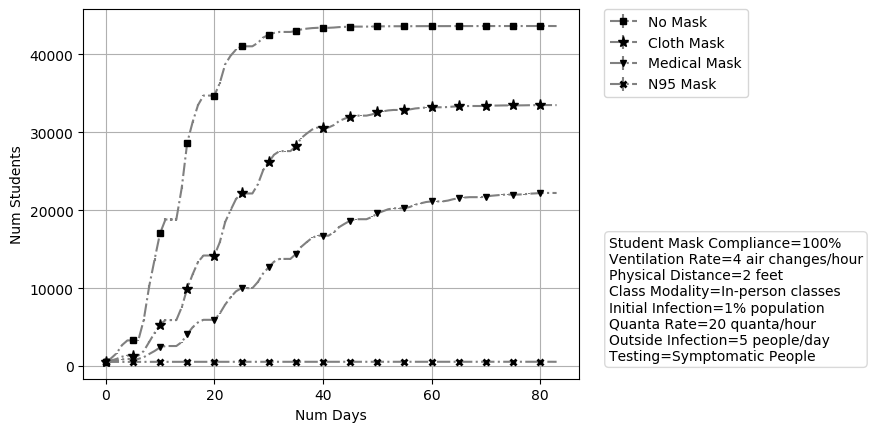}
\caption{Impact of different mask types on cumulative infected students due to the community transmission of COVID-19 within university campus}
\label{fig:individual-masking2}
\end{figure}

 Figure \ref{fig:individual-masking1} shows the impact of different Student Mask Compliance on the cumulative number of infected students due to community transmission within campus. A decrease of 23.25\% in cumulative infected students can be observed when there is a change from no student mask compliance to strict mask policy adherence. In Figure \ref{fig:individual-masking2}, we fix the student mask compliance to be 100\% and vary the mask types. We observe that the regular use of N95 masks by students will significantly reduce the spread. However, it's also expensive for them to use these masks every day. 


\subsubsection{Physical Distancing} 
\label{sec:Physical_distancing}

COVID-19 spreads among people in close proximity for sufficient time. In our simulation, we use physical distance as a radius to calculate the area of a circle which acts as a substitute for the area per student. This value is used to determine the room volume of a class based on the total number of students attending the class. The product of room volume and ventilation rate parameter (4 ac/h) is used as the room ventilation rate parameter in the Wells-Riley equation \eqref{eqn:wells-riley taylor approx}. 

\begin{figure}
\centering
\includegraphics[width=\textwidth]{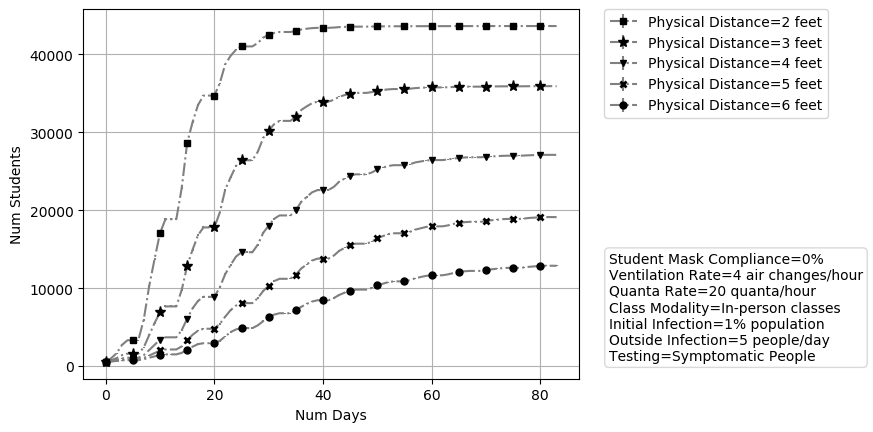}
\caption{Impact of different physical distancing on cumulative infected students due to the community transmission of COVID-19 within university campus}
\label{fig:individual-physicalDistancing}
\end{figure}

Center for Disease Control (CDC) recommends a physical distance of 6 feet \cite{socialdistancing}. To analyze physical distancing policy, we vary the physical distance from 2 (personal space \cite{hall1966hidden} in normal times) to 6 feet. We analyze the physical distancing dimension by assuming scheduling of all classes in-person, no adherence on wearing mask among students, and only symptomatic people are tested. 
As seen in Figure \ref{fig:individual-physicalDistancing}, an increase from 2 to 6 feet in physical distance can decrease the cumulative infected by 70.50\%, which suggests that it's crucial to avoid close contact with other people even though they are not showing any symptoms. 

\subsubsection{Class Modality}

\begin{figure}
\centering
\includegraphics[width=\textwidth]{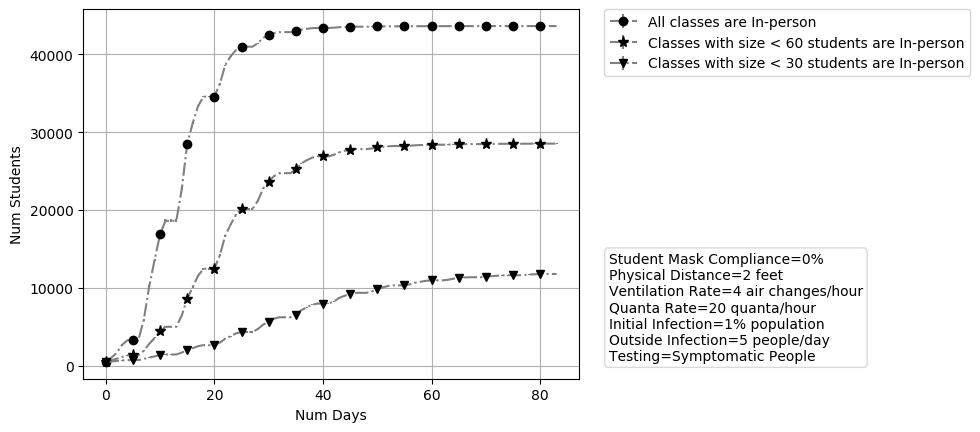}
\caption{Impact of different class size thresholds for class modality on cumulative infected students due to the community transmission of COVID-19 within university campus}
\label{fig:individual-classModality}
\end{figure}

Large gatherings, like classes, increase the likelihood of spreading the virus \cite{schuchat2020public, adam2020clustering}. Therefore, we analyze the impact of class modality by varying the maximum class size. We set the mask compliance among students to be 0\%, physical distancing is reduced to 2 feet, and only symptomatic infected people are tested. In particular, we vary the maximum in-person class sizes to 30, 60 or all in-person, as shown in Figure \ref{fig:individual-classModality}.

A change from all classes being in-person to restricting classes with more than 30 students to be online, can bring a decrease of 72.99\% in cumulative infected. Large classes act as hubs where typically students from various departments study together, which increases their potential exposure to other students. Consequently, it creates short paths between students of different departments (communities) \cite{weeden2020small}, which exacerbates the community spread. Therefore, avoiding large classes can be very effective in controlling the spread of the disease.

\subsubsection{Testing} \label{sec:individual-testing}
In the simulation, we can implement testing policies to detect infection spread in the network. On a given day, each person has a test state of tested positive, tested negative, or not tested. The testing capacity each day is assumed to be limited due to constraints from time, labor, money, manufacture of testing supplies, etc. Available tests are preferentially used to test symptomatic people. People who turn symptomatic go for testing the next day. Based on the testing results on each day, we then create a list of classes attended by the positively tested people, called the 'Contact Traced' (CT) classes. The rest of the classes come under 'Non-Contact Traced' (NCT) classes. We can implement various policies to sample students or instructors from CT and NCT classes differently for testing purposes. In our current simulation, the available tests are preferentially used to test symptomatic people. The rest of the tests are used to test students only in ’Contact Traced’ (CT) classes. Contact traced people are sampled based on the log of their infectability.

In the simulation, we use the test accuracy statistics from a publicly available Infection Testing manufactured by Inbios \footnote{ https://inbios.com/smart-detecttm-sars-cov-2-rrt-pcr-kit/}, with $96.7\%$ sensitivity and $100\%$ specificity. If an individual has a False Negative test result, he / she can still attend classes and spread the virus. If an individual has a False Positive test result, he / she will be quarantined in the simulation. We also define a 'testing gap day' parameter which stops people from getting tested within a particular time frame. In the simulation we set the 'testing gap day' as 3 days, because this is the average number of days after which an infected person becomes infectious. This value can be changed depending upon the availability of tests and other policies. In addition, we set the mask compliance among students to be 0\%, physical distancing to be 2 feet, and all classes to be held in-person in our simulation.

\begin{figure}
\centering
\includegraphics[width=\textwidth]{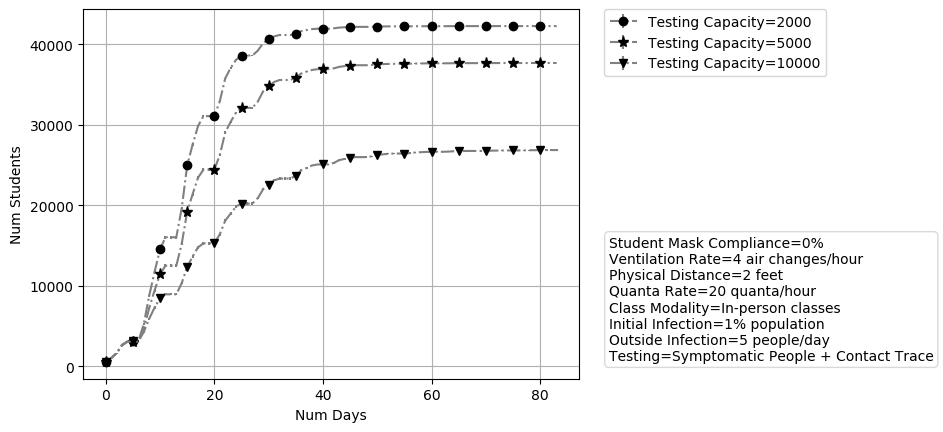}
\caption{Impact of different test capacities on cumulative infected students due to the community transmission of COVID-19 within university campus}
\label{fig:individual-testing}
\end{figure}

To study this policy dimension, we vary the test capacity by 2000, 5000 and 10,000 tests per day. As shown in Figure \ref{fig:individual-testing}, 5x testing capacity (from 2000 to 10000) can lead to 36.45\% decrease in cumulative infected, which is less effective as compared to other individual dimensions we studied.

\subsection{Discussion}

The higher education community finds itself in uncharted territory due to COVID-19, which has limited the functioning of the community. Operational policies being implemented also carry social and monetary costs, which are presently unclear, but will impact the future. Thus, implementing cost-effective policies with high impact on controlling the spread of the disease is extremely important. 

Masking and physical distancing are economically cheaper to implement. However, as shown in Figure \ref{fig:individual-masking1} and \ref{fig:individual-physicalDistancing}, when these policies are implemented in isolation,  71.57\% and 27.50\% student population respectively still gets infected. These two policies also have high social cost for implementation.
In March, when understanding of the pandemic was nascent, policies such as self-isolation and social distancing were recommended to flatten the curve. 
Feelings of loneliness and isolation that are exacerbated during social distancing have caused mental health to suffer and lead to  increased substance use, and elevated suicidal ideation \cite{czeisler2020mental, sher2020impact}. Masking as a voluntary policy would likely lead to insufficient compliance, would be perceived as less fair, and could intensify stigmatization such as negative emotional responses, social labeling, or prejudicial attitudes \cite{betsch2020social}. 

On the other hand, policies like mass testing and shifting classes with huge enrollments online bear the financial burden and may not be as effective as expected in isolation, unless all other recommendations are being strictly followed by the students. Scientists at the University of Illinois developed a quick, inexpensive saliva test and started doing 10,000 to 15,000 tests per day. However, mass testing and contact tracing alone doesn't guarantee a control over the spread, as shown in our simulation results in Section \ref{sec:individual-testing}, which also results in 58.3\% population being infected by the end of the semester. Students' defying quarantining/self-isolation when being testing positive has also led to another spike in infections \cite{nytparty}. 

These policies when implemented individually don't seem to be enough to control the disease spread, but they can work well in conjunction with each other. We present a case study of a combination of these policies in our next section. 



%% file: case_study.tex
Here we present a detailed case study of the impact of various decisions taken by the University of Minnesota, one of the largest universities in the country, as it went through the process of developing and implementing its Sunrise Plan \cite{umnsunrise}. On March 11th, 2020 the University of Minnesota (UMN) suspended in-person instruction, including field experiences and classes, across five campuses and are moved to online functioning. The assortment of all the decisions taken to re-open UMN for fall semester is called a Sunrise Plan.

\begin{figure}
\centering
\includegraphics[width=\textwidth]{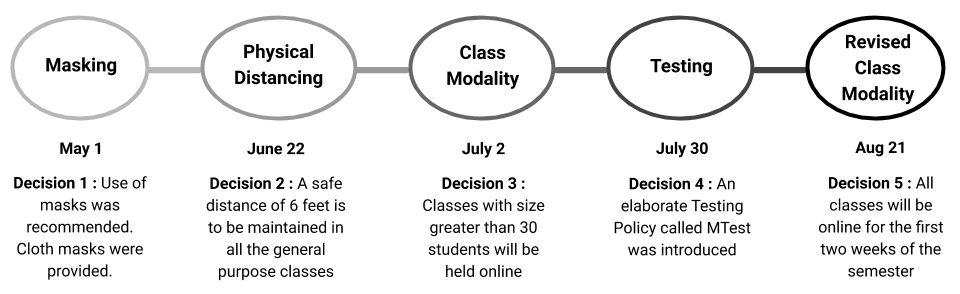}
\label{epidemiological}
\caption{Sunrise Plan decision timeline}
\label{fig:sunrise}
\end{figure}

The Sunrise Plan is summarized in Figure \ref{fig:sunrise}. Under UMN's Sunrise Plan, recommendations for the use of masks and social distancing were made as early as May 1. Cloth masks are being provided to all students and employees who are on campus and required in specific settings. As of June 22,  based on the physical distancing requirements outlined by the Minnesota Department of Health \cite{ihecovid19} and UMN's own medical and public health experts \footnote{https://safe-campus.umn.edu/sunrise-plan\#social-distancing}, decision on maintaining 6 feet physical distance in general-purpose classrooms was adopted. Course delivery modality was entered into the university course scheduling system by the end of July 2. It was also decided that any in-person class meetings will end right before Thanksgiving (Nov 26), and the remainder of the class meetings, including final exams, will be entirely online. Led by UMN's Health Emergency Response Office (HERO), a Testing and Tracing Advisory Team was formed and the team introduced an elaborate Testing Policy called MTest \footnote{https://safe-campus.umn.edu/return-campus/mtest-protocols-and-response} on July 30. On Aug 21, it was announced that all classes will be wholly online for at least first two weeks of the fall semester.

\begin{figure}
\centering
\includegraphics[width=0.99\textwidth]{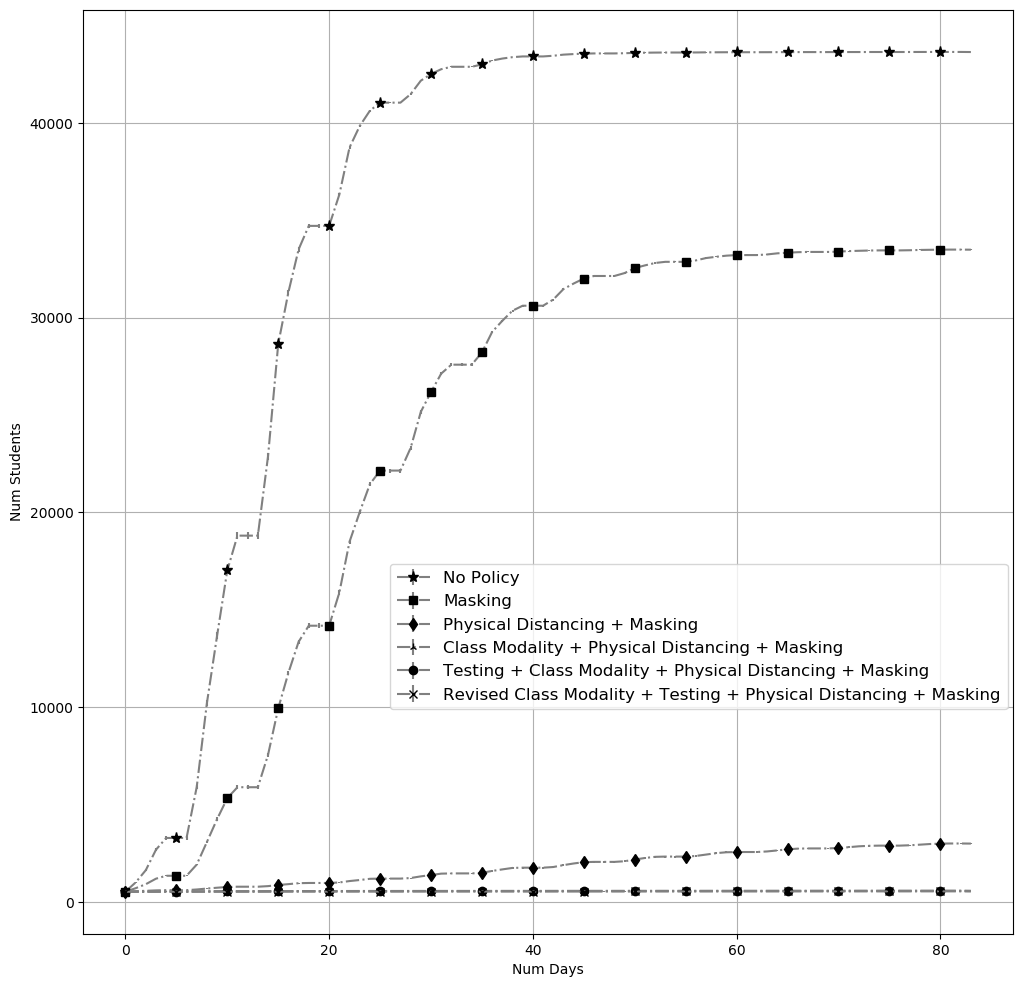}
\caption{Impact of various decisions on Cumulative Infected due to the Community transmission of COVID-19 within university campus, implemented by UMN over time. }
\label{fig:sunrise plan}
\end{figure}

As the university considered and implemented alternative policies over time, we modeled the impacts of those decisions, including requiring masks, the extent of physical distancing in classrooms, introduction of class modalities as to designate classes as in-person, blended, remote, or online, and testing protocols. This was done over a roughly 4-month period from May 1st to August 21st, 2020. As each decision was considered, the impact of various alternatives for it were evaluated. Since decisions were made one at a time, in evaluating the impact of downstream decisions, the system also incorporated the earlier decision made in the Sunrise Plan. In Table \ref{tab:cumulative_infected_weeks_sunrise} and Figure \ref{fig:sunrise plan}, we show the combined results of various policies adopted at different stages of Sunrise Plan. The results can be applied to get a measure of confidence for each decision. Our results show that, even just the proper adherence of masking and physical distancing can bring a drastic difference in the cumulative infected in comparison to individual policies.

\begin{table}
\centering
\begin{tabular}{ |p{3cm}||p{1.1cm}|p{1.1cm}|p{1.1cm}|p{1.1cm}|p{1.1cm}|p{1.1cm}|p{1.1cm}|p{1.1cm}| }
 \hline
 Policies & \multicolumn{8}{|c|}{Cumulative Infected at the end of following weeks} \\
 \hline
 & Week 1 & Week 2 & Week 3 & Week 4 & Week 6 & Week 8 & Week 10 & Week 12 \\
 \hline
 No Policy & 3291.36 & 18808.63 & 34707.19 & 41032.28 & 43404.45 & 43605.42 & 43627.34 & 43629.96\\
 \hline
 M   & 1350.66 & 5895.81 & 14185.09 & 22147.44 & 30609.55 & 32862.29 & 33370.08 & 33484.94 \\
 \hline
 PD + M & 612.29 & 789.27 & 981.02 & 1205.96 & 1766.62 & 2335.64 & 2753.98 & 3002.83\\
 \hline
 CM + PD + M & 542.76 & 558.7 & 566.93 & 570.98 & 575.68 & 578.98 & 581.88 & 579.99\\
 \hline
 T + CM + PD + M & 541.17 & 553.91 & 559.0 & 561.19 & 563.81 & 566.63 & 569.89 & 567.76\\
 \hline
 RCM + T + PD + M & 533.04 & 533.04 & 537.07 & 538.97 & 541.72 & 545.09 & 547.86 & 545.5 \\
 \hline
\end{tabular}
\caption{Impact of various decisions on Cumulative Infected due to the Community transmission of COVID-19 within university campus, implemented by UMN over time. M=Masking, PD=Physical Distancing, CM=Class Modality, T=Testing, and RCM=Revised Class Modality.}
\label{tab:cumulative_infected_weeks_sunrise}
\end{table}


%% file: conclusion.tex
COVID-19 has presented administrators of higher education institutions with a completely new problem, i.e. {\it what are the right decisions to make to for operating an educational campus during the pandemic, such that educational objectives can be met, while ensuring health safety for the entire community}. The history on this problem is short, i.e. only around 6 months, since the last pandemic with this level of societal impact, i.e. the Spanish flu of 1918, happened at a time when higher education was not much of an integral part of the society.

This paper is among the first set of efforts to address this problem. Our approach is unique in that it presents a flexible simulation system that contains a suite of model libraries, one for each major system component. The simulation system merges agent based modeling (ABM) and stochastic network approach, and models the interactions of individual entities, e.g. students, instructors, classrooms, residences, etc. in great detail. For each decision to be made by administrators, the system can be used to predict the impact of various choices, and thus enable the administrator to make a suitable decision.

A detailed case study of the University of Minnesota's Sunrise Plan \cite{umnsunrise} was presented. Specifically, as various decisions were made in sequence, their impact was assessed using this system. The results were used to get a measure of confidence in each decision. We believe this flexible tool can be a valuable asset for various kinds of organizations to assess their infection risks in pandemic-time operations.

{\bf Acknowledgement:} The authors acknowledge Krishnamurthy Iyer, Kelly Searle, and Rachel Croson for their constructive feedback at various stages of this research.